\documentclass[twocolumn]{aastex631}

\usepackage{newtxtext,newtxmath}
\usepackage{float}
\usepackage{natbib}
\usepackage[T1]{fontenc}
\usepackage{amsmath}

\DeclareMathOperator*{\argmin}{arg\,min}

\usepackage{graphicx}	% Including figure files
\usepackage{amsmath}	% Advanced maths commands
\usepackage{hyperref}
\hypersetup{
    colorlinks=true,
    linkcolor=blue,
    filecolor=magenta,      
    urlcolor=cyan,
}
\usepackage[toc,page]{appendix}
\shorttitle{A low-rank approach to CCD image defringing}
\shortauthors{Prunet}
\submitjournal{PASP}
\accepted{October 27, 2021}
\begin{document}
% Keep the title short and informative.
\title{A low-rank approach to image defringing}

% The list of authors, and the short list which is used in the headers.
% If you need two or more lines of authors, add an extra line using \newauthor

\author{Simon Prunet}

% List of institutions
\affiliation{Université Côte d'Azur, Observatoire de la Côte d'Azur, CNRS, Laboratoire Lagrange, Bd de l'Observatoire, CS 34229, 06304 Nice cedex 4, France}
\affiliation{Canada-France-Hawaii Telescope Corporation, 65-1238 Mamaloahoa Hwy, Kamuela, HI 96743, USA}
\affiliation{CNRS and Sorbonne Université, UMR 7095, Institut d’Astrophysique de Paris, 98 bis Boulevard Arago, F-75014 Paris, France.}
% These dates will be filled out by the publisher
%%%\date{\today}

% Enter the current year, for the copyright statements etc.
%%\pubyear{2015}

% Don't change these lines
%%%\begin{document}
%%%\label{firstpage}
%%%\pagerange{\pageref{firstpage}--\pageref{lastpage}}

\begin{abstract}
In this work, we revisit the problem of interference fringe patterns
in CCD chips occurring in near-infrared bands due to multiple light
reflections within the chip. We briefly discuss the traditional approaches
that were developed to remove these patterns from science images,
and mention their limitations. We then introduce a new
method to globally estimate the fringe patterns in a collection of
science images without additional external data, allowing for some variation 
of the patterns between images. We demonstrate this new method on near-infrared images
taken by the CFHT wide-field imager Megacam.
\end{abstract}

\section{Formulation of the problem, and past solutions}

CCD images in the red (e.g i' and z') bands are subject to a specific
systematic effect called "fringing". Emission lines from molecules or free radicals
in the upper atmosphere create interference patterns within the chips. 
Indeed, due to their reduced quantum efficiency in these bands, 
old CCD chips models act like Fabry-Pérot interferometers
for these quasi-monochromatic atmospheric emissions. This results
in sinusoidal patterns superimposed on the science images. The details
of the patterns are however complicated by the surface irregularities of
each chip, which depend on their manufacturing process. 
This systematic effect is additive, and its amplitude is proportional 
to the brightness of the sky lines emission.

One way to estimate fringe patterns is to produce uniform calibration exposures,
with line emissions lamps that reproduce the sky emission with sufficient
accuracy: this was for instance implemented at the Mayall Telescope with a neon
calibration light \citep{Howell2012}. The quality of the fringe pattern
obtained in that way is however limited by the spectral energy distribution difference
between the calibration light and the sky emission. Another way to produce a fringe 
template is to take
a median of (adequately rescaled) science images, taken at different
positions on the sky (thus mitigating the impact of astrophysical objects
in the images), and do some kind of robust regression of this pattern
on individual science images \citep[see e.g.][]{Valdes1998,Snodgrass+Carry2013},
based on suitably located control pairs of pixels. This is the spirit
of the method that is currently implemented in the Megacam elixir
pipeline at CFHT \citep{Magnier2004}.

Both methods assume that the fringe pattern is perfectly stable, and thus
do not take into account the time evolution of the pattern, probably sourced by small variations 
of the atmospheric line emission ratios. One approach to address this problem \citep[Elixir-LSB, e.g.][]{Ferrarese2012} 
is to observe sequences of images with sufficient dithering scales,
during a short period of time (typically one hour) over which the atmospheric
variability (and thus the changes in fringe pattern morphology) is expected to be small. 
This approach has been used with success to remove fringe patterns, as well as the large scale scattered
light pattern on Megacam images due to internal reflections in the
optical system.

In this work, we present a new method to remove fringe patterns together
with camera-related, large-scale diffuse light patterns from a set
of Megacam images, not necessarily taken in a sequence. This method
allows for intrinsic variations of the fringe patterns, and relies on the assumption
that only a few, properly weighted templates suffice to explain
the fringe pattern variations. It is very similar in spirit to the
well known Principal Components Analysis (PCA) method.
The software developed in this context is publicly available \href{https://github.com/simon-prunet/defringe}{here}.
We note that a very similar approach to defringing has been independently developed by \cite[][Appendix A]{Bernstein2017}; their clever use of image decimation allows their method to be used on full DECam focal plane images.

\subsection{Link to the PCA method}

Let us first slightly simplify the problem, and assume that the science
images taken at night contain no astrophysical sources (unreasonable!) 
but only the (possibly varying) fringe patterns, plus Gaussian
i.i.d noise (a reasonable assumption for background-dominated images
once properly re-scaled by their mean value). Let us in addition assume
that we know \emph{in advance} the maximum number $k$ of templates
we need to explain the variations of the fringe pattern. Organizing
the data into a large $(n_{\rm pix},n_{\rm obs})$ matrix $D$ where each chip image
is flattened into a long column vector, the problem of fringe pattern estimation
can be written in the following way:
\[
\hat{F}=\argmin_{F}||D-F||_{F}\quad{\rm such}\,{\rm that}\quad{\rm rank}(F)\lesssim k
\]

The solution to this problem is well known, it is the projection of
$D$ onto its first $k$ leading left singular vectors \citep[Eckart-Young theorem][]{Eckart1936}:
\[
\hat{F}=UH_{k}(\Sigma)V^{T}
\]

where $D=U\Sigma V^{T}$ is the singular value decomposition of $D$,
and $H_{k}$is the hard-thresholding operator which keeps only the
first $k$ singular values of $\Sigma$. The $k$ first principal
components correspond to the first $k$ columns of the matrix of left-singular
vectors $U$. This is the essence of the principal component analysis.

\subsection{Formulation of the problem}

In reality, there are several differences between the fringe pattern estimation
and the ideal problem mentioned above, where a straight PCA analysis
would give the answer. First of all, the effective number of modes
needed to explain the fringe pattern variations is not known in advance,
and secondly the presence of astrophysical sources in the images makes
the assumption of simple i.i.d Gaussian noise inappropriate at the
location of the sources. Is the last problem serious? Meaning, are
the principal components robust to the presence of outliers in the
images? Unfortunately they are not. PCA suffers from the same lack
of robustness as classical linear regression, and one strong outlier
is enough to corrupt the PCA results.

To address the problem of the presence of astrophysical sources in the images, we
will use a masking strategy, where pixels contaminated by astrophysical
sources will be removed from the cost function. The problem of estimating
the masks will be addressed later in this document. We will also replace
the constraint on the rank of the solution by a \emph{proxy} which
happens to have nice mathematical properties 
\citep[tightest convex relaxation of the rank functional][]{Fazel2002,Candes2009,Recht2010} 
with a penalization term proportional to the nuclear norm of $F$, since we don't know
in advance the exact number of modes will be needed. The convexity of the cost
function implies the existence of a global minimum, as well as efficient
minimization algorithms even if the penalization is not differentiable everywhere.
The problem we propose to solve thus reads:
\begin{equation}
\hat{F}=\argmin_{F}\mu||F||_{*}+\frac{1}{2}||P_{\Omega}(D-F)||_{F}^{2}\label{eq:cost-function}
\end{equation}

where $P_{\Omega}$ is a projection operator that only keeps non-masked
pixels, and $||F||_{*}$ is the nuclear norm of the matrix $F$, i.e. 
the sum of its singular values. The parameter $\mu,$ which controls
the relative weight of the $\chi^{2}$ term and the regularization
term, must be chosen in such a way that singular values dominated
by noise are strongly suppressed, while those dominated by signal
(fringe patterns) are preserved (usual bias-variance compromise).
It can be shown \citep{Candes2009} that a good choice is $\mu=(\sqrt{n_{\rm pix}}+\sqrt{n_{\rm obs}})\sqrt{p}\sigma$,
where $p$ is the fraction of observed pixels, and $\sigma$ is the
noise standard deviation. 

Equation~\ref{eq:cost-function}, without the projection operator,
has a known solution in terms of the SVD decomposition of $D=U\Sigma V^{T}$:
\begin{equation}
\argmin_{F}\mu||F||_{*}+\frac{1}{2}||D-F||_{F}^{2}=US_{\mu}(\Sigma)V^{T}\label{eq:soft-thresholding}
\end{equation}
where $S_{\mu}(\Sigma)={\rm diag(max(\sigma_{1}-\mu,0)},...,max(\sigma_{n_{\rm obs}}-\mu,0))$
is the so-called soft-thresholding operator with parameter 
$\mu$\footnote{Note that this is the proximal operator to the nuclear norm}.
The difficulty thus comes from the presence of source pixel masks,
and solving Eq.~\ref{eq:cost-function} is a low-rank matrix completion
problem. This type of problems has been recently heavily studied due
to its numerous applications, and several algorithms to solve the
problem have been proposed. 

We have retained two different algorithms, one for its simple formulation
and the second one for its ability to handle possible variations of
the problem formulation. They both show similar performance when applied
to the defringing problem. The corresponding algorithms are described
in Appendix A. The first class of algorithms is based on iterative resolutions of the
soft thresholding problem of
Eq.~\ref{eq:soft-thresholding}  with the missing entries of the data matrix filled with
the solution obtained at the precedent iteration. This method is dubbed SOFT INPUTE by their
authors \citep{Mazumder2010}, due to its relation to the soft thresholding step. The second 
algorithm by \cite{Shen2010} is an accelerated proximal gradient (APG) method, which also relies on
soft thresholding iterations.

Nuclear norm regularization, and the associated soft thresholding of singular values,
leads to well-behaved algorithms due to its convex nature, but also induces a bias in 
the recovered singular values of the fringes modes. To alleviate this, several possible
methods could be used. \cite{Mazumder2010} mention this problem, and propose to use their 
HARD INPUTE alternative algorithm after first approaching the desired solution with SOFT INPUTE.
This is a viable approach, with the caveat that the alternative algorithm, based on hard thresholding
(which applies to rank functional minimization) is not convex. Another alternative is to use a 
"truncated" nuclear norm, with soft-thresholding limited to the $n_{\rm obs}-k$ last singular values.
This approach, advocated by \cite{Hu2013}, also leads to a non convex optimization problem (albeit better
behaved). 
To avoid these problems altogether, we decided to proceed in two steps:
\begin{itemize}
    \item First, use a convex optimization algorithm (either SOFT INPUTE or APG, see appendix~\ref{sec:appendix}),
    \item In a second time, select the relevant $k$ first left singular vectors, and compute their respective
    coefficients in the different images using a regular linear regression on the valid pixels
\end{itemize}
In this approach, the fringe modes are not co-optimized together with their coefficients in each image till
the end, but conversely we only deal with convex optimization sub-problems. In practice it gives excellent results,
as we will show in the next section.

\section{Application to Megacam z-band images}

In this section, we will describe the application of the proposed low-rank method to images taken
by the Canada-France-Hawaii Telescope (CFHT) Megacam imager. 
This instrument\footnote{\url{https://www.cfht.hawaii.edu/Instruments/Imaging/Megacam/}} is a wide
field imager with 40 e2V CCD42-90 chips, with electromagnetic wide-band filters similar
to those of the Sloan Digital Sky Survey \citep{Gunn1998}.

\subsection{Astrophysical source masks\label{subsec:Astrophysical-source-masks}}

The first problem that needs to be addressed to apply the former method
to astrophysical images is the design of the pixel masks for all images.
To this end, we adopted the following scheme: first we compute a median
fringe pattern (in a very similar way to the CFHT elixir pipeline), by first
removing from each image its median value (assumed to be dominated
by the sky background emission), and re-scale them by their exposure
time. We thus compute, for each pixel, the median of the $n_{\rm obs}$ re-scaled
images to create a median fringe pattern. 

We then do some kind of robust regression of this median fringe pattern in
each image to get a first, approximate set of defringed images. Finally,
a simple $\kappa.\sigma$ clipping is applied to the images to create
the source masks, where $\sigma$ is obtained with a robust
Median Absolute Deviation estimator. After some experimentation, a value of $\kappa=2$
was adopted, as a compromise between masking bright sources and creating
too many holes in the images. We then optionally enlarge the masks
around each source, to remove the signal in the tails of the detected
sources.

This simple $\kappa.\sigma$ clipping procedure could of course be
replaced by a more sophisticated source extraction software such as
SExtractor \citep{Bertin1996}. The impact of the details of
the source extraction method on the defringing process remains to be investigated
in details.

\subsection{Defringing results }

\begin{figure*}
\begin{centering}
\includegraphics[width=1\textwidth]{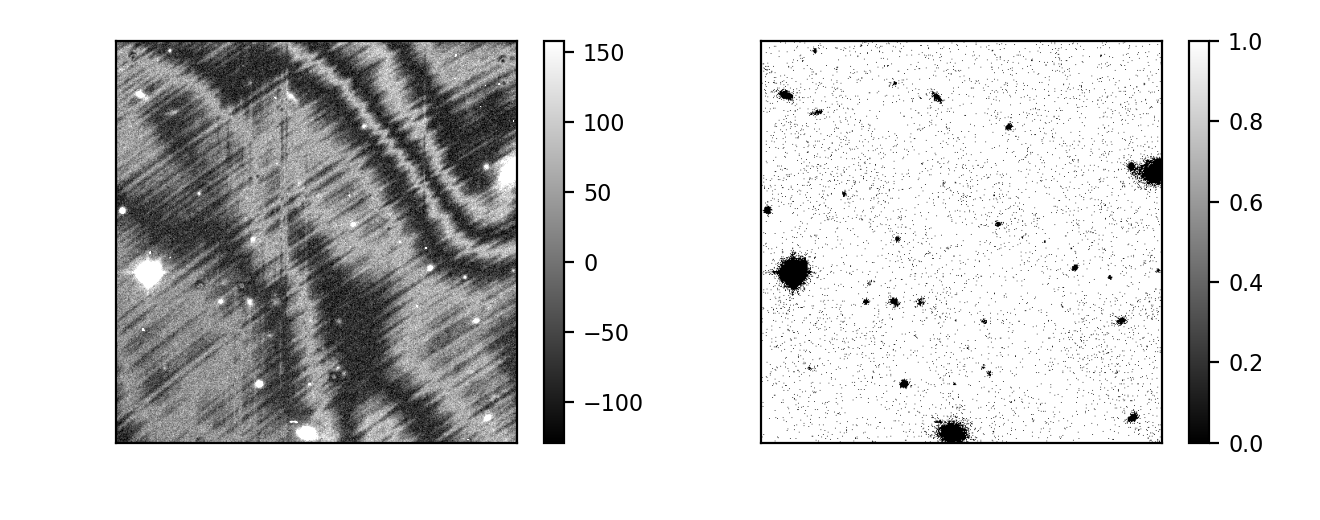}
\par\end{centering}
\caption{Left panel shows a $500\times500$ pixels sub-image of CCD chip \#13 of a typical
z-band MegaCam image. Fringes are clearly visible on this image. Right
panel: corresponding source mask obtained following the procedure
described in Section~\ref{subsec:Astrophysical-source-masks}.\label{fig:fringes}}
\end{figure*}

As a proof of concept, we applied the method to a $(500\times500)$ pixels subset of a given CCD
chip, for $n_{\rm obs}=37$, z-band images taken from a single MegaCam observing run. At the blue edge of this
filter (around $830$ nm), the quantum efficiency of the e2V chips is at most of $40\%$ and rapidly decreasing, leading
to strong fringing effects. The images
come in several consecutive sequences separated by typically a few
nights. Figure~\ref{fig:fringes} shows a typical bias subtracted,
flat-fielded sub-image of a given CCD chip, with the corresponding
mask where bright astrophysical sources, as well as cosmic ray hits have been flagged.

\begin{figure*}
\begin{centering}
\includegraphics[width=1\textwidth]{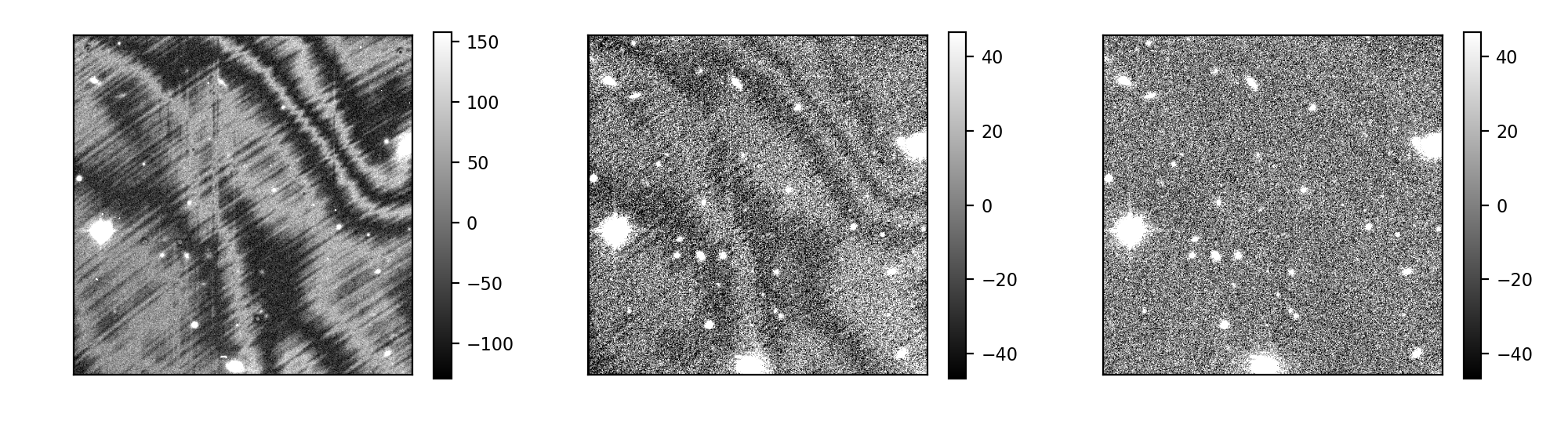}
\par\end{centering}
\caption{\emph{Left panel} shows the same, original $500\times500$ pixels image,
where the fringe pattern is clearly visible.
\emph{Middle panel} shows the median-subtracted fringe-removed image
produced by the elixir package in production at CFHT. \emph{Right
panel} shows the result of the method described in this note. The residual contamination is
clearly much smaller than in the middle panel, this underlines the importance of taking
into account the time evolution of the fringe pattern over several nights. \label{fig:defringed}}

\end{figure*}

Figure~\ref{fig:defringed} shows the result of removing the estimated
fringe pattern from the same image, using either our new method or
the CFHT elixir pipeline recipe, based on a single median fringe template. 

\begin{figure*}
\begin{centering}
\includegraphics[width=1\textwidth]{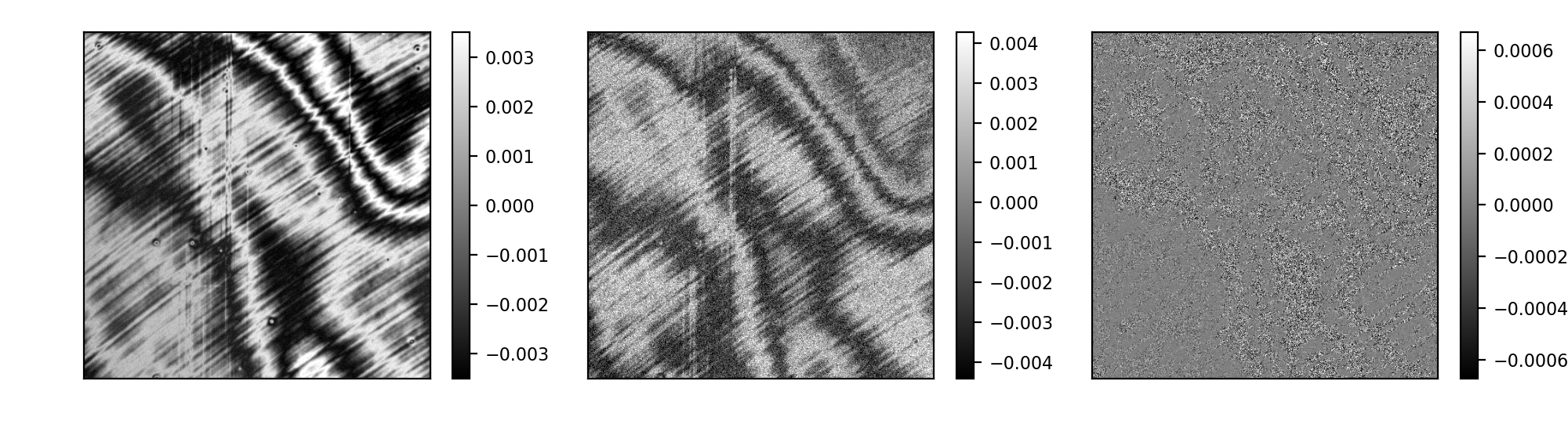}
\par\end{centering}
\caption{From left to right: projection of the low-rank fit to the image (Eq.~\ref{eq:cost-function}),
onto its first three singular vectors. The first two images look similar
(ignoring their amplitude and relative noise contamination differences),
except for tiny translations/deformations, which is expected from
small variations of the OH emission line ratios \citep[see][]{Howell2012}. 
The third singular vector is already heavily noise dominated, 
which shows that the effective rank of the fringe pattern
matrix for this sequence of images is equal to $2$. \label{fig:modes}}

\end{figure*}

Finally, Figure~\ref{fig:modes} shows the first three left-singular
vectors of the low-rank fit (solution of Eq.~\ref{eq:cost-function}).
Here, only two modes have a reasonable SNR, and indeed the SOFT INPUTE algorithm
only assigns significant singular values to the first two modes.
For the latter, singular values of $5184$ and $168$
ADUs respectively, are produced by SOFT INPUTE, increasing to $5689$ and $695$ ADUs after linear regression on the modes. 

Peak-to-peak estimates in the first and second images of Figure~\ref{fig:defringed}are the order
of 150 and 20 ADUs respectively. Keeping only the first mode would
thus result in fringe residual contamination at the level of 20 ADUs
peak-to-peak in this particular image.

\begin{figure}
    \begin{centering}
    \includegraphics[width=1\columnwidth]{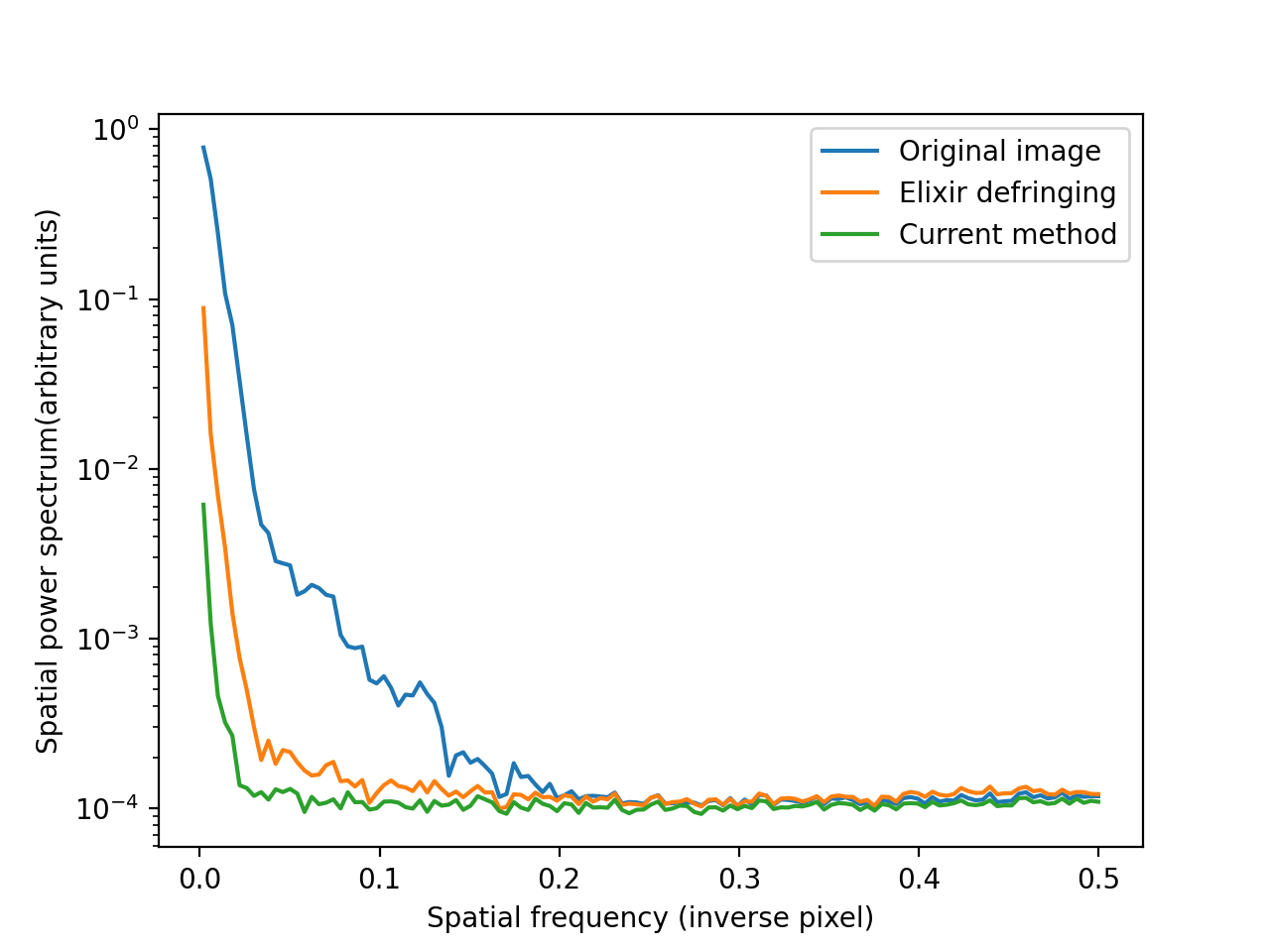}
    \par\end{centering}
    \caption{Spatial power spectra of resp. the original image (blue line), the median-subtracted image produced by the elixir pipeline (orange line) and the image produced by the current method (green line). Astronomical objects were masked out in the analysis, and edge correction was performed to decouple the different spectral modes. One can see at least a ten-fold improvement in the residual contamination on the largest scales with the current method, with a power level compatible with photon noise up to scales of 50 pixels or larger.}
    \label{fig:powerspectra}
\end{figure}

To quantify the contamination levels in the different images of Figure~\ref{fig:defringed}, we have computed the spatial power spectra of the images, masking out the astrophysical objects that would otherwise dominate the statistics. Decoupling of the power spectra was obtained with the Namaster software \citep{Alonso2019}, using its flat sky version of the Master algorithm \citep{Hivon2002}.
These power spectra are shown in Figure~\ref{fig:powerspectra}, where it appears that the residual contamination in the images processed with the current approach is reduced by a factor of at least ten on the largest scales, compared to the median template regression of the CFHT elixir pipeline.

\begin{figure}
\begin{centering}
\includegraphics[width=1\columnwidth]{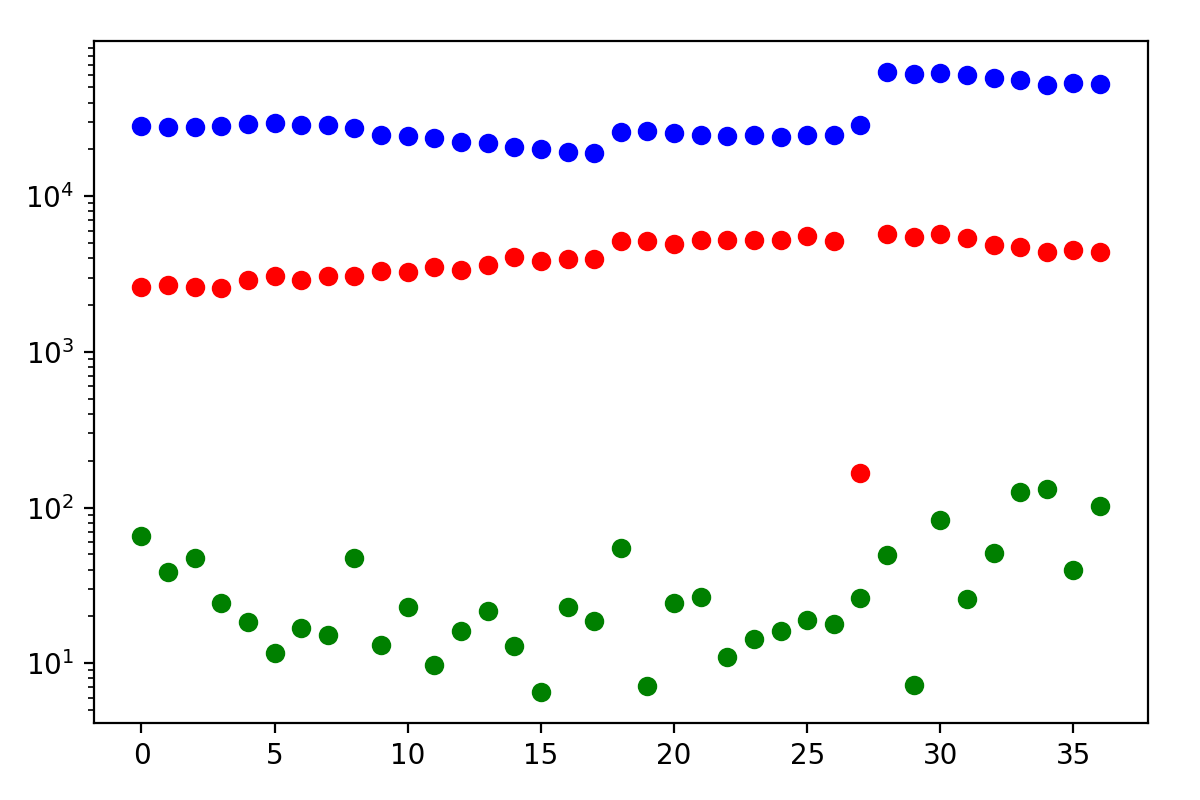}
\par\end{centering}
\caption{Decomposition coefficients of the low-rank fits onto the three principal
(left-singular) modes. Blue, red and green symbols correspond respectively
to the first, second and third modes. The observations can be grouped into
three main groups in this plot, that correspond to three different
nights of observation. The evolution of the fringe patterns is evident
in this plot. \label{fig:mode-weights}}

\end{figure}

The right-singular modes (corresponding to the three largest
singular values), multiplied by the corresponding singular values,
are shown in Fig.~\ref{fig:mode-weights}, in blue, red and green
symbols respectively, for the 37 images considered here. These are
the relative weights of the (left-singular) modes in the low-rank
fit to each observed image. The observing time distributions appear
clearly in this figure: three distinct groups of images, corresponding
to three different nights, can be seen. Finally, the coefficients of the 
third left-singular mode appear to have low amplitude and vary substantially
between images: they are dominated by noise, and reflect again the fact that the
effective number of fringe modes in this data set is equal to $2$.

\section{Discussion and Conclusion}

The method presented in this work, based on the assumption that the individual
fringe patterns in different images of an observing run can be decomposed into
a linear combination of a small number of modes, has been shown to be very successful
at estimating (and thus removing) the fringe patterns from real images of the CFHT
instrument Megacam. In particular, this method learns the modes from the data itself,
without needing any external calibration data, in a way very similar to Principal Component
Analysis. In order to make the analysis robust to the presence of bright sources and cosmic ray hits,
a masking procedure has been used, and only data outside those bright regions are used.
Finally, the method relies on two convex optimization steps, which ensures the existence and unicity
of a global minimum, and allows the use of iterative methods of known convergence properties.

In practice, the SOFT-INPUTE method, with the regularization coefficient $\mu$ chosen as prescribed by
\cite{Candes2009}, converges to machine precision with about ten iterations (for a total of twenty including
the warm start phase). However, each iteration is potentially quite costly, since it consists in a Singular
Value Decomposition of a very tall matrix of size $n_{\rm pix},n_{\rm obs}$. For the sub-images considered here,
$n_{\rm pix}=250000$, and $n_{\rm obs}=37$ images, a multi-threaded, CPU-based SVD takes around $2$ seconds on a Xeon $2.10$ GHz server with 40 cores.
For a full e2V CCD with $n_{pix}=9445376$, the wall time goes up to $40$ seconds per SVD. However, GPU acceleration can be
very useful here: using a Tesla V100 GPU card and the CUBLAS library, these wall times are respectively brought down to
$90$ ms and $2$ s for the sub-image and the entire CCD image. This makes it a viable method, even for the very large Megacam focal plane, as long as CCD chips are processed independently.

Also, the number of modes needed to adjust the fringe patterns is typically very small (in our data set, only two modes were
necessary to adjust the fringe patterns), therefore only partial SVDs are needed, opening the possibility to use iterative Lanczos like
methods \citep{Masana2018}, polar decompositions and projections on the 2-norm ball \citep{Cai2013}, or randomized SVD methods \citep{Halko2011}.

Finally, this method is readily applicable to other CCD imagers suffering from similar fringe contamination patterns, as well as large scale, slowly changing scattered light patterns. We note however that this method is based on a linear, additive model of the contamination, which in principle makes it ill-suited for the fringe patterns caused by similar interference effects in the flat field modes of spectrographs, as in the latter case the fringe patterns are of a multiplicative nature (the interference is sourced by the light of the astrophysical object and not primarily by the emission from the sky). 

However, it is possible that after a first multiplicative correction by an average flat field, residual corrections could be further modelled by an additive, linear model, in which case the present method could be generalized to spectral data. Such fringing also happens in the case of infrared detectors, and \cite{Argyriou2020} noticed that the fringing patterns, in the context of the JWST MRS spectrograph, are modulated by the spatial extension of the source being observed, which further complicates the problem. Such a generalization will be investigated in a future work.

\begin{acknowledgements}
I would like to warmly thank An-Yi Bu for her help during her internship at CFHT, Pascal Fouqué, Claire Moutou, Kanoa Withington, 
Jean-Charles Cuillandre and Gary Bernstein for enlightening discussions, Colin Snodgrass for the link to the ESO Faint Object Camera archive images on which preliminary versions of the code were tested, 
and André Ferrari for the suggestion to replace the singular value thresholding step with a neural network based denoising step. This will be investigated
in a future work.
I would like to thank the Canada-France-
Hawaii Telescope (CFHT) which is operated by the Na-
tional Research Council (NRC) of Canada, the Institut
National des Sciences de l'Univers of the Centre National
de la Recherche Scientifique (CNRS) of France, and the
University of Hawaii. The observations at the CFHT
were performed with care and respect from the summit
of Maunakea which is a significant cultural and historic
site.
I was supported during this work by a BQR grant from Laboratoire Lagrange (UCA,OCA,CNRS).
\end{acknowledgements}

\appendix
\section{Two algorithms for the cost function minimization with
nuclear norm}\label{sec:appendix}

Here we will give some details on the two algorithms that we tried
to solve the minimization problem of Eq.~\ref{eq:cost-function}.
The first one, dubbed SOFT-INPUTE, is taken from \cite{Mazumder2010}. Its
formulation is very simple, and its nice convergence properties come
from using as ``warm start'' the solution of the minimization problem
with a larger value of the penalization weight $\mu$. The algorithm
thus uses a decreasing sequence of penalization weights $\mu_{1}>\dots>\mu_{K}=\mu(\sigma)=(\sqrt{n_{\rm pix}}+\sqrt{n_{\rm obs}})\sqrt{p}\sigma$:
\begin{enumerate}
\item Initialize $F^{{\rm old}}=0$
\item Do for $k=1,\dots,K$:

\begin{enumerate}
\item Repeat:

\begin{enumerate}
\item Compute the (partial) SVD of the current completion: $USV^{T}=P_{\Omega}(D)+P_{\Omega^{\perp}}(F^{{\rm old}})$
\item Compute $F^{{\rm new}}=US_{\mu_{k}}(\Sigma)V^{T}$, soft-thresholding
of the current completion
\item If $||F^{{\rm new}}-F^{{\rm old}}||_{F}^{2}/||F^{{\rm old}}||_{F}^{2}<\epsilon$
exit
\item $F^{{\rm old}}=F^{{\rm new}}$
\end{enumerate}
\item $\hat{F}_{\mu_{k}}=F^{{\rm new}}$
\end{enumerate}
\item Keep $\hat{F}_{\mu(\sigma)}$as the solution, or keep the entire sequence
as warm starts for the HARD-INPUTE algorithm.
\end{enumerate}
The HARD-INPUTE algorithm minimizes the following non-convex cost
function:
\[
\hat{F}_{H}=\argmin_{F}\frac{1}{2}||P_{\Omega}(D-F)||_{F}^{2}+\mu\times{\rm rank}(F),
\]
where the algorithm is similar to that of SOFT-INPUTE, but replacing
the soft-thresholding in (ii) above with the hard thresholding operation,
keeping unchanged all singular values larger than $\mu_{k}$, and
putting the others to zero. Instead of initializing $F^{{\rm old}}=0$,
there is a large gain in using the output of SOFT-INPUTE, i.e. to
start with $F^{{\rm old}}=\hat{F}_{\mu_{k}}$. Note that for HARD-INPUTE,
if the target penalization weight $\mu$ is known (e.g. given by some
heuristic relationship with the noise level as in $\mu(\sigma)$),
then one can directly start at $k=K$ and skip the iterations with
larger values of $\mu$, using $\hat{F}_{\mu(\sigma)}$as the initial
value. An alternative to using HARD-INPUTE is to select the few left singular
modes, computed by SOFT-INPUTE, that correspond to non-zero singular values,
and regress their respective coefficients in the different images.
This last sub-problem is convex and therefore the existence of a unique, global
minimum is guaranteed: we followed this approach in practice to bypass possible convergence
issues linked to the non-convexity of HARD-INPUTE.

The second algorithm to minimize the cost function of Eq.~\ref{eq:cost-function}
that we tested is called the Accelerated Proximal Gradient method
\citep{Shen2010}. This algorithm is more general than SOFT-INPUTE,
in the sense that it can be applied to the more general convex minimization
problem:
\begin{equation}
\min_{F}L(F)=\mu g(F)+f(F)\label{eq:APG-cost}
\end{equation}
 where $f$ is differentiable everywhere and has a Lipschitz continuous
gradient, whereas $g$ is convex but not necessarily differentiable.
This method is interesting if a fast solution for the following minimization
sub-problem is available:
\begin{equation}
\min_{F}\mu g(F)+C||F-G||_{F}^{2}\label{eq:APG-sub}
\end{equation}
where $C$ is a constant. Let us consider the Taylor expansion of
the function $f$ at second order around point $Y$, because of its
Lipschitz continuous gradient it will be bounded by:
\begin{eqnarray*}
Q(F,Y) & \equiv & f(Y)+\langle\nabla f(Y),F-Y\rangle+\frac{\tau}{2}||F-Y||_{F}^{2}\\
 & = & \frac{\tau}{2}||F-G||_{F}^{2}+{\rm cste}
\end{eqnarray*}
where the constant term is only a function of $Y$, and $G=Y-\tau^{-1}\nabla f(Y)$.
So for a given $Y$, we replace the minimization problem of Eq.~\ref{eq:cost-function}
by
\[
\min_{F}\mu g(F)+\frac{\tau}{2}||F-G||_{F}^{2}
\]
which is precisely the sub-problem of Eq.~\ref{eq:APG-sub}. We see
that by appropriately choosing the update scheme of $Y$ and $F$ we
can iterate on the sub-problem minimizations to get a descent algorithm
for the original problem of Eq.~\ref{eq:APG-cost}. It was shown
(e.g. Bertsekas 1999) that the following algorithm has an $\mathcal{O}(1/k^{2})$
convergence rate:
\begin{enumerate}
\item Choose $F^{0}=F^{-1}$ in the definition domain of $g,$ $t^{0}=t^{-1}\in[1,\infty)$.
\item For $k=0,1,\dots$ do:

\begin{enumerate}
\item $Y^{k}=F^{k}+(t^{k-1}-1)/t^{k}(F^{k}-F^{k-1})$
\item $G^{k}=Y^{k}-\tau^{-1}\nabla f(Y^{k})$; and compute $F^{k+1}=\argmin_{F}\mu g(F)+\frac{\tau}{2}||F-G^{k}||_{F}^{2}$
\item $t^{k+1}=\frac{1+\sqrt{1+4(t^{k})^{2}}}{2}$
\end{enumerate}
\end{enumerate}
with $\tau$ the Lipschitz constant of $\nabla f$. In the case of
the matrix completion problem with nuclear norm regularization of
Eq.~\ref{eq:cost-function}, we can identify $g(F)\equiv||F||_{*}$
and $f(F)=\frac{1}{2}||P_{\Omega}(F-D)||_{F}^{2}$, so that $G^{k}=Y^{k}-\tau^{-1}P_{\Omega}(F-D)$
because $P_{\Omega}$ is a projection operator. For the same reason,
we get $\tau=1$ here. The sub-problem of Eq.~\ref{eq:APG-sub}
is solved by the soft-thresholding solution $F^{k+1}=US_{\mu_{k}/\tau}(\Sigma)V^{T}$where
$G^{k}=U\Sigma V^{T}$. This algorithm can be accelerated by dynamically
changing the parameter $\mu$, by choosing a decreasing sequence that
eventually reaches the target regularization $\mu(\sigma)$: $\mu_{k}=\max(\kappa.\mu_{k-1},\mu(\sigma))$,
where the update can be made every n cycles, or whenever the rate
of change of $F^{k}$ becomes slow. This sequence of decreasing $\mu_{k}$
is very similar to the ``warm start'' sequence of SOFT-INPUTE.

%%%% BIBLIO
\bibliography{authors}
\bibliographystyle{aasjournal}

\end{document}